\documentclass[11pt,a4paper]{article}
\pdfoutput=1

\usepackage{graphicx}
\usepackage{caption}
\usepackage{subcaption}
\usepackage{bm}
\usepackage{amsmath}
\usepackage{amssymb,xfrac}
\usepackage[bbgreekl]{mathbbol}
\usepackage{cite,color,float}

\def\be{\begin{equation}}
\def\ee{\end{equation}}
\def\bea{\begin{eqnarray}}
\def\eea{\end{eqnarray}}

\begin{document}

\pagestyle{plain}

\begin{center}
~

\vspace{1cm} {\large \textbf{
Electric Charge in Presence of Fixed \\
Monopole-Antimonopole Pair
}}

\vspace{1cm}

Niloufar Barghi-Janyar ~~~~~ and ~~~~~ Amir H. Fatollahi$\,^*$

\vspace{.5cm}

{\it Department of Physics, Alzahra University, \\ P. O. Box 19938, Tehran 91167, Iran}

\vspace{.3cm}

$^*\,$\texttt{fath@alzahra.ac.ir}

\vskip .8 cm
\end{center}

\begin{abstract}
The dynamics of an electric charge $e$ in presence of a fixed monopole
pair $\pm g$ is considered. Depending on the ratio of 
the angular momentum to $e\,g/c$, the effective potential 
may consist a minimum valley between the poles or a finite depth well. 
In the classical limit the charge may be bound to the poles inside the minimum 
valley or well. In the quantum theory, due to the finite barrier between
the minimum and the potential at infinity, only approximate or quasi bound-states
are possible. Based on the Rayleigh-Ritz variational method the energy 
and eigen-functions are obtained. The implications on the proposed confining mechanism 
based on the dual picture of the so-called Meissner effect is pointed. 
\end{abstract}

\vspace{1cm}



\newpage

\section{Introduction}

The magnetic field by a single magnetic monopole of strength $g$ at origin is given by
\begin{align}\label{1}
\mathbf{B}=g\, \mathbf{r}/ r^3
\end{align}
This can be obtained by means of
$\mathbf{B}=\bm{\nabla}\times \mathbf{A}$ with
\cite{zwa}
\begin{align}\label{2}
\mathbf{A}=g \frac{\mathbf{r}\times \mathbf{\hat{n}}}{r(r-\mathbf{r}\cdot\mathbf{\hat{n}})}
\end{align}
Except along the half-line $\mathbf{r}=\alpha\,\mathbf{\hat{n}}$ with $\alpha\geq 0$, the 
above vector potential is regular \cite{zwa}. In fact the singular line acts as an
intensive field inside the ultra-thin solenoid to cancel the outward flow by 
the monopole; the so-called Dirac string \cite{jackson}. By the choice $\mathbf{\hat{n}}=\mathbf{\hat{z}}$ (\ref{2}) reduces to the well 
known expression \cite{sakurai}
\begin{align}\label{3}
\mathbf{A}=g\frac{1+\cos\theta}{r\,\sin\theta}~ \bm{\hat{\phi}}
\end{align}
Demanding the wave-function of the electric charge $e$ in presence of the monopole
to be single-valued, the quantization condition follows\cite{dirac,milton}:
\begin{align}\label{4}
e\, g /c= q\,\hbar/2, ~~~~~~~~~~q=0,\pm 1,\pm2,\cdots
\end{align}
It is known that a monopole and an electric charge do not make bound-state
\cite{tamm}. The bound-state problem of two dyons (states with both electric
and magnetic charges) is considered in \cite{zwa}, where  
the exact energy eigen-values are obtained for cases that the dyonic bound-states are possible. 

In the present work the problem of the dynamics of an 
electric charge in presence of a fixed monopole-antimonople
pair is considered. In the Hamiltonian formulation of the 
system the effective potential in the 
$\rho z$-plane describes the nature of possible
bound-states. Accordingly it is found that, depending on 
the ratio of angular momentum and the combination of charges in (\ref{4}),
the effective potential may consist minimum valley or well
accompanied by a saddle-point, or no extrema at all. 
In the classical regime, when the potential 
has a minimum the bound motion of the electric charge is possible. 
In the quantum theory, as the minimum is 
apart from the tail $V\to 0$ at infinity with finite 
height barrier, the formation of true bound-state is not possible. 
It is known that, due to the tunneling effect through a barrier of
finite height and width, only the so-called 
quasi bound-states are possible. However, for the case with large barrier height these
states may treated approximately as bound-states. 

The results by the present study can directly be used in the dual picture, 
in which the dynamics of a monopole is considered in the 
presence of two fixed opposite electric charges. 
Apart from pure theoretical curiosity, the problem in 
dual picture can shed light on the confinement mechanism in gauge theories based 
on the dual picture of the so-called Meissner effect in type-II superconductors. 
It is known that the electric charges make
a circulating motion around the thin magnetic fluxes inside the superconductors, 
hence the name of vortex for these thin magnetic fluxes. 
In fact the circulating motion of electric charges around the 
magnetic fluxes prevent the magnetic fields to spread
inside the superconductor, leading to the confinement of 
the hypothetical monopole and anti-monopole pair inserted inside the sample. 
In the proposed confinement mechanism, it is the motion of the monopoles that prevents the 
spreading of the electric fluxes, leading to the confining phase of the theory \cite{nambu,mand,thooft}.

The organization of the rest of paper is as follows. In Sec.~2 the setup of 
the system is presented together with the Hamiltonian formulation of the 
classical system. The nature of dynamics is discussed based on the effective
potential in the $\rho z$-plane, together with the bound-state and 
the scattering problems. In Sec.~3 the quantum dynamics is formulated and 
developed. Based on the variational Rayleigh-Ritz method the energy eigen-values 
are obtained for the cases that the approximate bound-states are expected. 
In Sec.~4 the concluding remarks and discussions are presented.

\section{Classical Dynamics}
The magnetic field by a monopole $+g$ and antimonopole $-g$ on 
$z$-axis, respectively at $z=\ell$ and $z=-\ell$, is given by:
\begin{align}\label{5}
\mathbf{B}= g  \left(
\frac{\mathbf{r}_+}{r_+^3} -\frac{\mathbf{r}_-}{r_-^3}
\right)
\end{align}
in which (see Fig.~1)
\begin{align}\label{6}
\mathbf{r}_\pm &= \mathbf{r} \mp \, \ell\, \mathbf{\hat{z}}\cr
&=\rho \, \bm{\hat{\rho}} + (z\mp\ell)\,\mathbf{\hat{z}}\cr
r_\pm&=\sqrt{\rho^2+(z\mp \ell)^2} 
\end{align}

\begin{figure}[t]
	\begin{center}
		\includegraphics[scale=0.9]{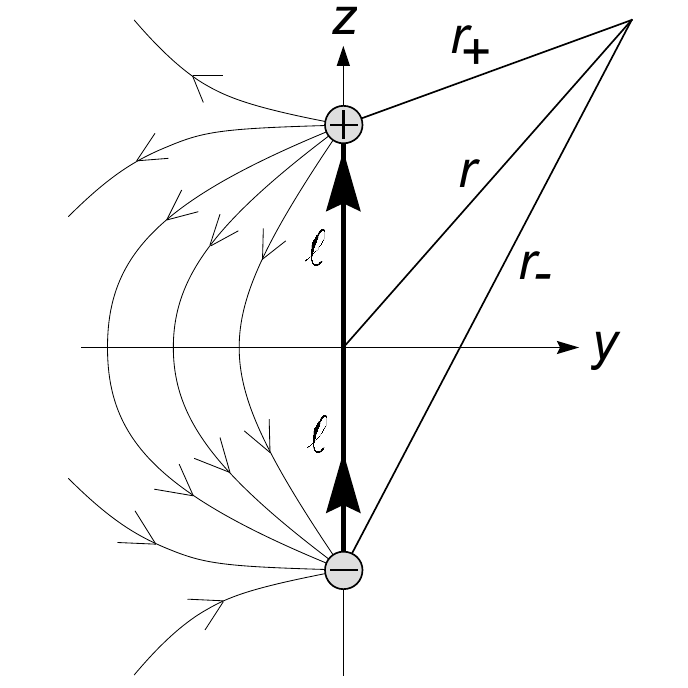}
	\end{center}
	\caption{\small The setup with monopole pairs. 
The only singular part is the segment between the poles. The  thick arrow exhibits the direction of intense field flux inside the Dirac string.}
\end{figure}

The Lorentz equation of motion of the elctric charge $e$ with 
mass $\mu$,  $\mu\,\ddot{\mathbf{r}}= e\, \dot{\mathbf{r}}\times \mathbf{B} /c$,
in the cylindrical coordinates $(\rho,\phi,z)$, leads to 
\begin{align}\label{7}
\mu\,(\ddot{\rho}-\rho\,\dot{\phi}^2) &= \frac{eg}{ c} \rho \dot{\phi} 
\left( \frac{z-\ell}{r_+^3 } -\frac{z+\ell}{r_-^3}\right)
\\\label{8}
\mu\,(\rho\,\ddot{\phi}+2\dot{\rho}\,\dot{\phi}) &=\frac{eg}{c}
\left( \frac{\rho \dot{z} -\dot{\rho}(z-\ell)}{r_+^3}-
\frac{\rho \dot{z} -\dot{\rho}(z+\ell)}{r_-^3}\right)
\\\label{9}
\mu\,\ddot{z} &= \frac{eg}{ c}\rho^2 \dot{\phi} 
\left( \frac{1}{r_+^3 } -\frac{1}{r_-^3}\right)
\end{align}
Among others, one possible solution corresponds to
motion along the curve with almost fixed $\phi$ defined by 
\begin{align}\label{10}
\mathbf{v} \times \mathbf{B} \approx 0 ~~\to~~
\frac{dz}{d\rho}\approx \frac{B_z}{B_\rho}
\end{align}
It is easy to check that the uniform circular motion in the 
$xy$-plane around the $z$-axis also satisfies the above equations:
\begin{align}\label{11}
\rho\equiv \rho_0,~~~~~~~\phi=\omega\, t,~~~~~~~~~~  z\equiv 0
\end{align}
provided that
\begin{align}\label{12}
\omega = \frac{2\,e\, g\,\ell}{ \mu c\,  (\rho_0^2+\ell^2)^{3/2}}
\end{align}
Simple force analysis shows that stable circular motion against small perturbations can be expected. 
Later in the Hamiltonian formulation the condition for the stable solutions corresponding to two solutions in above are discussed in detail. 

\subsection{Effective Potential}

The Lagrangian formulation of the problem is possible. By the polar angles $\theta_\pm$ for monopoles (see Fig.~1), with the choice (\ref{3}) and
using $r_-\sin\theta_-=r_+\sin\theta_+=\rho$, we have 
\begin{align}
\mathbf{A}&=g\left(
\frac{1+\cos\theta_-}{r_- \sin\theta_-} -\frac{1+\cos\theta_+}{r_+ \sin\theta_+} \right)
\bm{\hat{\phi}}\cr
\label{13}
&=\frac{g}{\rho}~ (\cos\theta_- - \cos\theta_+) \,\bm{\hat{\phi}}
\end{align}
The form chosen in above is such that
the only singular part is on $z$-axis in $-\ell \leq z \leq \ell$. 
The Lagrangian
$ L=\frac{1}{2}\mu \,\dot{\mathbf{r}}^2 +\frac{e}{c}\, \dot{\mathbf{r}}\cdot\mathbf{A}$, using $v_\phi=\rho\,\dot{\phi}$,  then comes to the form
\begin{align}\label{14}
L=\frac{1}{2}\mu(\dot{\rho}^2+\rho^2\dot{\phi}^2 +\dot{z}^2) 
+\frac{eg}{c} \dot{\phi}\, (\cos\theta_- - \cos\theta_+)
\end{align}
The canonical momenta by the above are then
\begin{align}\label{15}
p_\rho&=\mu\,\dot{\rho},~~~~~~~~~~~~~~p_z=\mu\,\dot{z}, \cr
p_\phi&=\mu\,\rho^2\,\dot{\phi} 
+\frac{eg}{c}\, (\cos\theta_- - \cos\theta_+)
\end{align}
As the $\phi$ coordinate is absent in the Lagrangian its 
canonical momentum $p_\phi$ is conserved. The Hamiltonian of the 
system is easily obtained to be 
\begin{align}\label{16}
H=\frac{p_\rho^2+p_z^2}{2\,\mu} +  V_\mathrm{eff}(\rho,z)
\end{align}
in which the effective potential in the $\rho z$-plane is
\begin{align}\label{17}
V_\mathrm{eff}(\rho,z) &= \frac{1}{2\,\mu} \big[F(\rho,z) \big]^2
\end{align}
with 
\begin{align} \label{18}
F(\rho,z)& =
\frac{1}{\rho}\left(p_\phi - \frac{eg}{c} 
\,(\cos\theta_- - \cos\theta_+)\right)
\\\label{19}
&=\frac{1}{\rho}\left(p_\phi - \frac{eg}{c} 
\,\left(\frac{z+\ell}{r_-} - \frac{z-\ell}{r_+}\right)\right)
\end{align}
The last expression presents the explicit $\rho$ and $z$ dependences.
Evidently $V_\mathrm{eff} \geq 0$, with the zero as the 
absolute minimum. The extrema of the potential are obtained by either conditions:
\begin{align}\label{20}
\mathrm{Cond.~(1)}:&~~~~~~~~~~~~~~~~~F(\rho,z)=0,
\\\label{21}
\mathrm{Cond.~(2)}:&~~~~~
\partial_\rho F(\rho,z)=0,~~~~\partial_z F(\rho,z)=0
\end{align}
The first condition defines the zero valley $V=0$ along the below 
curve in the $\rho z$-plane:
\begin{align}\label{22}
\cos\theta_- - \cos\theta_+=\frac{z+\ell}{r_-} - \frac{z-\ell}{r_+}=\frac{c\, p_\phi}{eg}
\end{align}
It is easy to check that differentiating $d/dt$ of above leads to (\ref{10}), and so 
(\ref{22}) corresponds to the motion along the curve between two poles 
with fixed $\phi$ coordinate. 
Using the fact that $~0 \leq (\cos\theta_- - \cos\theta_+) \leq 2~$, 
the first condition is satisfied only when
\begin{align}\label{23}
0 \leq \frac{c\, p_\phi}{eg} \leq 2
\end{align}

The second condition (\ref{21}) defines the other types of extrema, not necessarily 
a minimum one. The condition (\ref{21}) leads to $z=0$ together with (see Appendix)
\begin{align}\label{24}
\frac{c\, p_\phi}{eg}=2\,\cos\theta_0 (1+\sin^2\theta_0),
~~~~~~~~~~~\rho_0 = \ell\, \tan\theta_0
\end{align}
by which the equilibrium distance $\rho_0$ in the $z=0$ plane 
is given. The plot of the right-hand side of (\ref{24}) is presented in 
Fig.~2. The above has solution for $0\leq \theta_0 \leq\pi/2$  when 
\begin{align}\label{25}
0 \leq \frac{c\, p_\phi}{eg} \leq \frac{8\sqrt{2}}{3\sqrt{3}}\simeq 2.18
\end{align}

\begin{figure}[t]
	\begin{center}
		\includegraphics[scale=0.6]{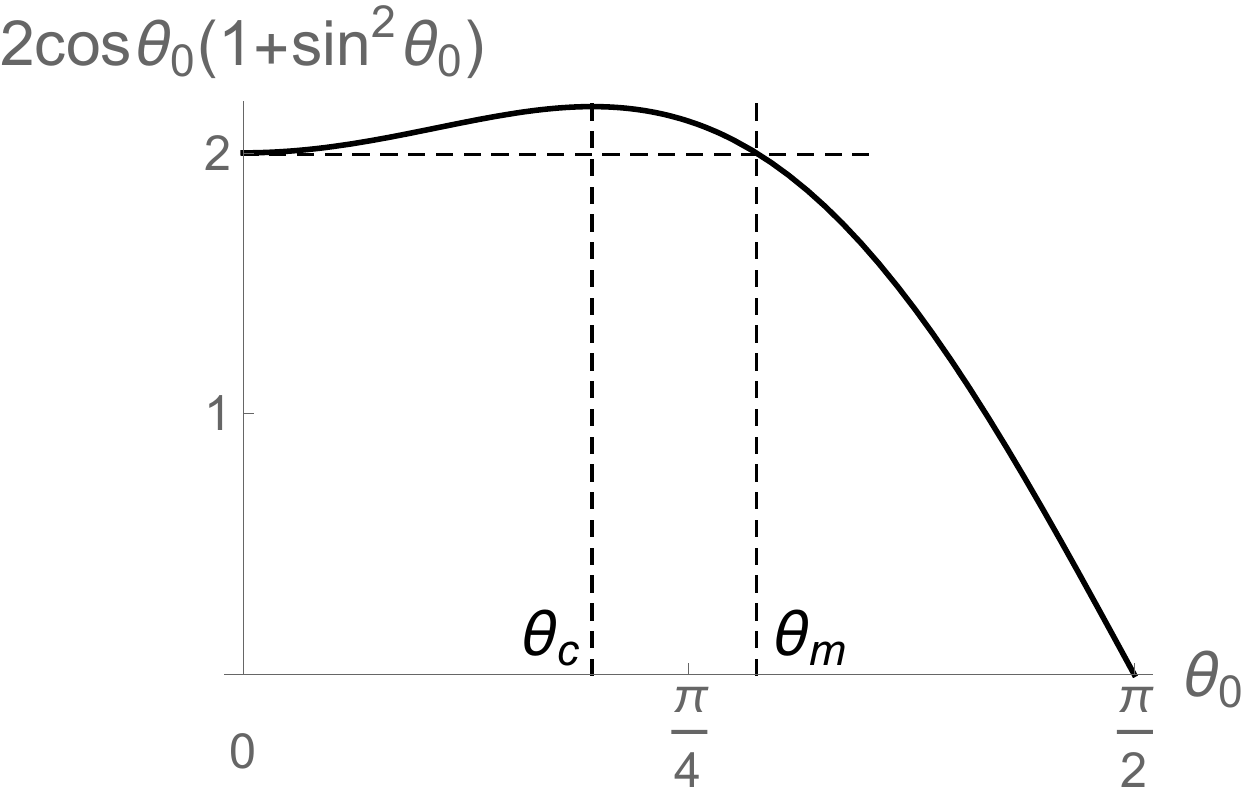}
	\end{center}
	\caption{\small The plot of right-hand side of (\ref{24}), with $\theta_c=\sin^{-1}\!\!\frac{1}{\sqrt{3}}$ and $\theta_m=\cos^{-1}\!\!\left(\frac{-1+\sqrt{5}}{2}\right)$.}
\end{figure}

The second derivative determines the nature of the extrema, by 
defining $\rho_c=\ell/\sqrt{2}$, for which we have
\begin{align}\label{26}
\partial_\rho^2 F &=\frac{4\,eg\,\ell}{c\,r_0^5\,\rho_0} \, (\rho_c^2-\rho_0^2) 
\\\label{27}
\partial_z^2 F &=\frac{6\,eg\,\ell}{c\,r_0^5} \, \rho_0,~~~~~~~~~~
\partial_\rho \partial_z F = 0 
\end{align}
By above for $\rho_0< \rho_c$ we have a minimum, otherwise it is a saddle-point
extremum (note $\partial_z^2 F >0$). 
The case with $\displaystyle{c\,p_\phi/(eg)=2}$ is special. In this case the extrema occur at 
\begin{align}\label{28}
\mathrm{minimum~at~\theta_0=0}:&~~~\rho_0=0,~~~~~~~-\ell \leq z \leq \ell 
\\\label{29}
\mathrm{saddle\!\!-\!\!point~at~\theta_0=\theta_m}:&~~~
\rho_0=\sqrt{\frac{2}{\sqrt{5}-1}}\,\ell,~~~~ z=0
\end{align}
with $\theta_m=\cos^{-1}\frac{-1+\sqrt{5}}{2}\simeq0.9046~\mathrm{rad}$ (see Appendix).
In summary, we have the following about the 
extrema of the effective potential:
\begin{itemize}
\item $0 \leq \displaystyle{\frac{c\, p_\phi}{eg}} \leq 2$: minimum valley along curve $F(\rho,z)=0$ between two poles accompanied by an outer saddle-point $\rho_0 >\rho_c$ and $z=0$ by the single solution of (\ref{24}) (the solution with $\theta_0 >\theta_m$ in Fig~2). A sample plot
of potential is presented in  Fig.~3a.
\item $2 \leq \displaystyle{\frac{c\, p_\phi}{eg}} \leq \displaystyle{\frac{8\sqrt{2}}{3\sqrt{3}}}$: 
an inner local minimum well at $\rho_{01}<\rho_c$ and $z=0$ accompanied by an 
outer saddle-point at $\rho_{02} >\rho_c$ and $z=0$ both by the double solutions of (\ref{24}) (two solutions with $0<\theta_0 <\theta_m$  in Fig~2). A sample plot of potential is presented in 
Fig.~3b.  The motion with constant $\rho_0$ is then a circular one with the constant angular 
velocity 
\begin{align}\label{30}
\dot{\phi} = \frac{1}{\mu\, \rho_0^2} \left( p_\phi - 
2\,\frac{eg\, \ell}{c\, r_0}\right) = 2\,\frac{eg\,\ell}{\mu\,c\,r_0^3}
\end{align}
with $r_0=\sqrt{\rho_0^2+\ell^2}$, which is 
exactly (\ref{12}) by the equations of motion.
\item $\displaystyle{\frac{c\, p_\phi}{eg}} \geq \displaystyle{\frac{8\sqrt{2}}{3\sqrt{3}}}$:
no local extrema
\end{itemize}
In the classical dynamics we can have stable motions about the minima.
As examples in Fig.~4 slightly perturbed motions about the two cases in above 
are presented. However as will see shortly, the situation is different in the quantum theory, in which a barrier with finite-width can not have 
real bound-states due to the tunneling effect. 

\vskip 1cm
\begin{figure}[H]
\centering
\begin{subfigure}{.5\textwidth}
\centering
		\includegraphics[width=.9\linewidth]{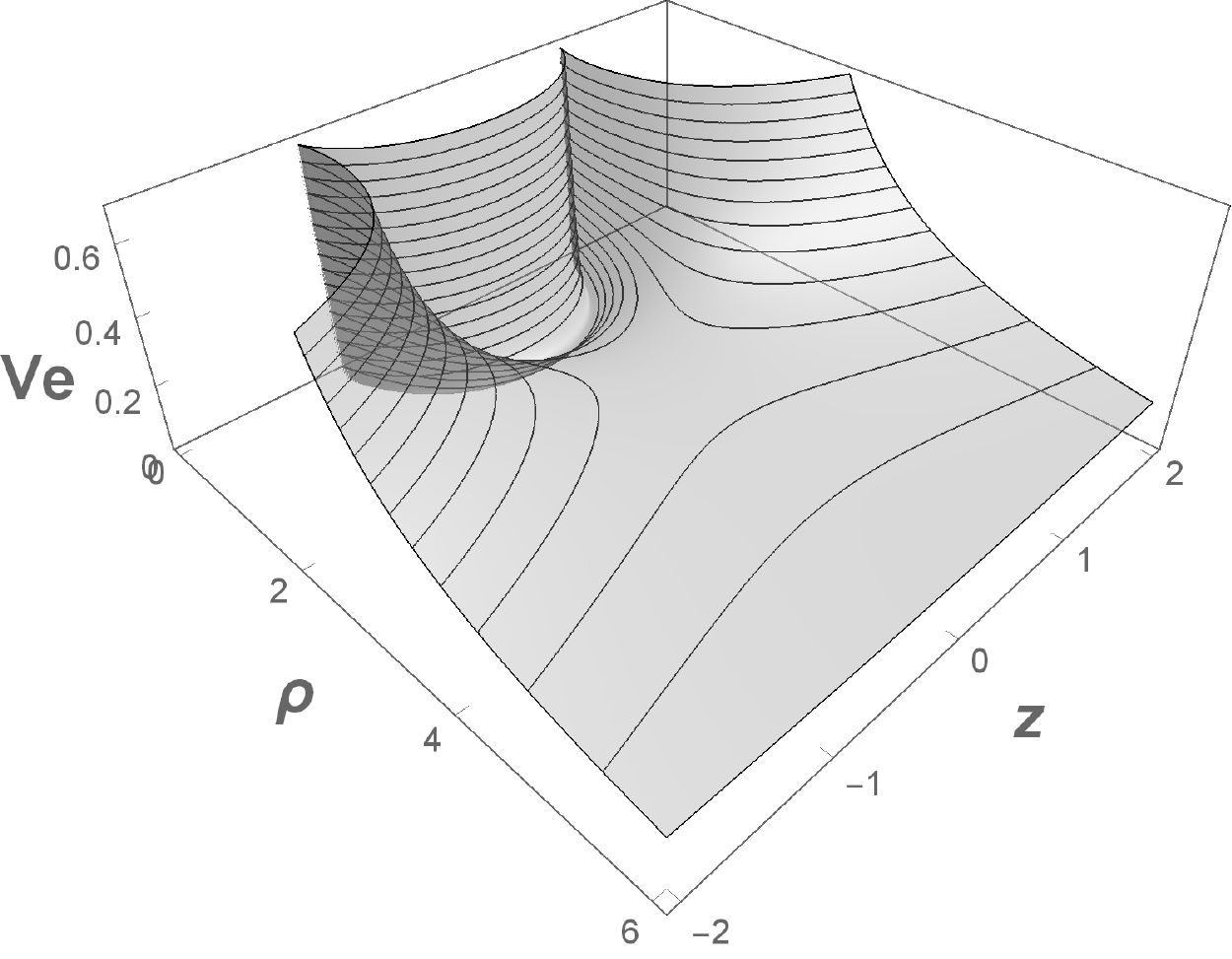}
\caption{\small $p_\phi=3$,  $eg/c=2.4$ and $\ell=1$}
\end{subfigure}%
\begin{subfigure}{.5\textwidth}
\centering
		\includegraphics[width=1.1\linewidth]{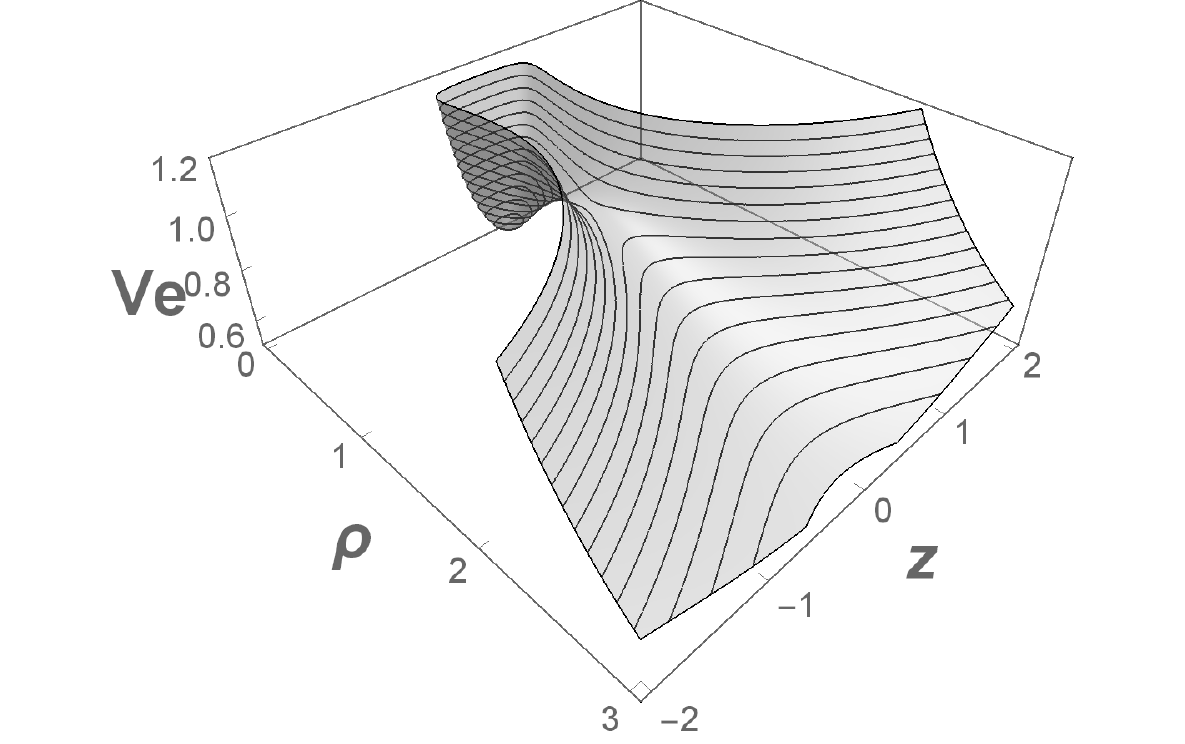}
\caption{\small $p_\phi=3$, $eg/c=1.44$, and $\ell=1$}
\end{subfigure}
	\caption{}
\end{figure}

\begin{figure}[H]
\centering
\begin{subfigure}{.5\textwidth}
\centering
		\includegraphics[width=.95\linewidth]{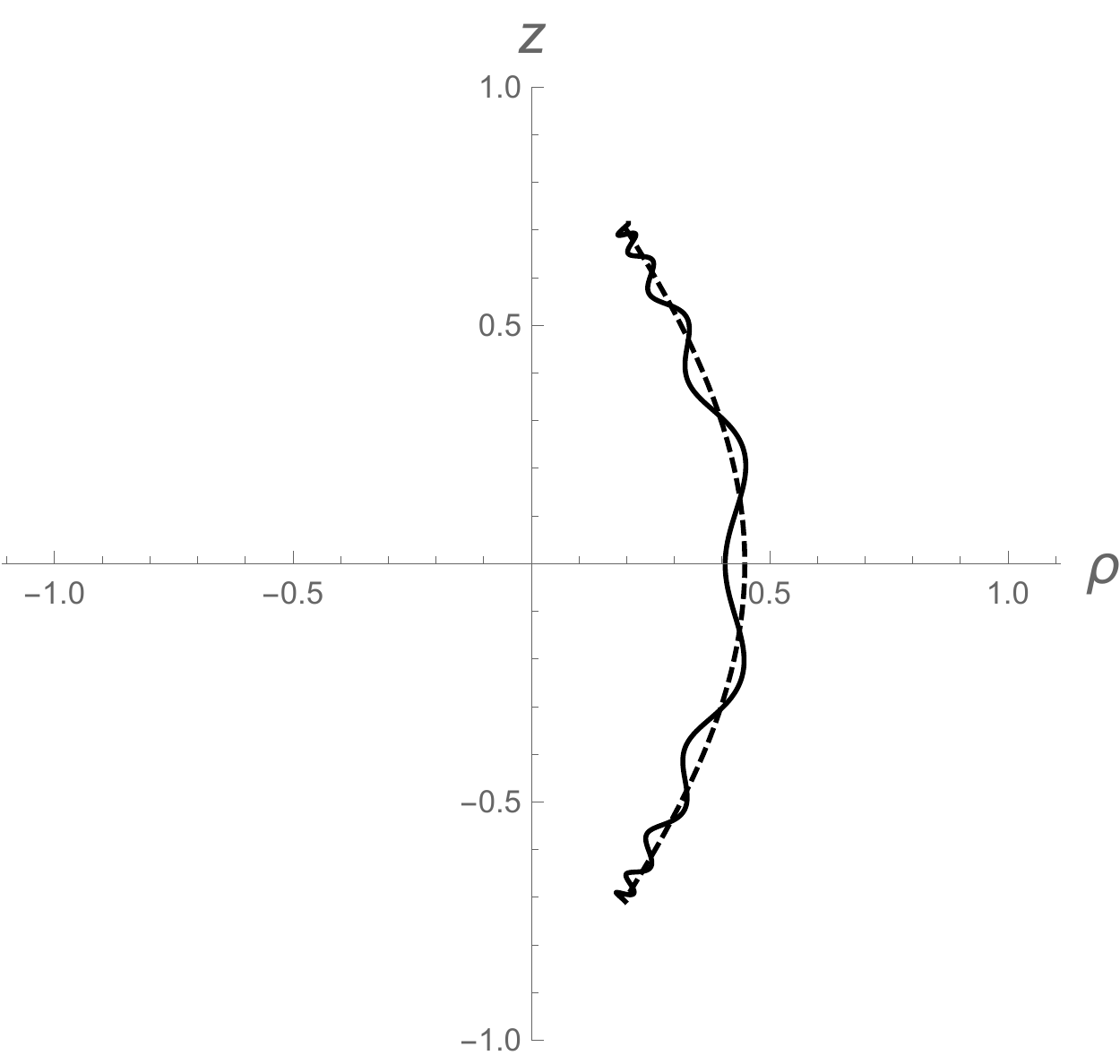}
\caption{\small The $\rho z$-projection of the 3D helical \\ motion 
(solid-line) around the curve\\ $F(\rho,z)=0$ (dashed-line).
The plot is by\\ values: $\mu=1$, $\ell=1$, $eg/c=2.8$,\\ $c\, p_\phi/eg=1.825$, $\rho_0=0.204$,
$\dot{\rho}_0=0.0204$, \\ $z_0=0.714$, $\dot{z}_0=-0.142$, $\phi_0=0$, $\dot{\phi}_0=5$.}
\end{subfigure}%
\begin{subfigure}{.5\textwidth}
\centering
		\includegraphics[width=.95\linewidth]{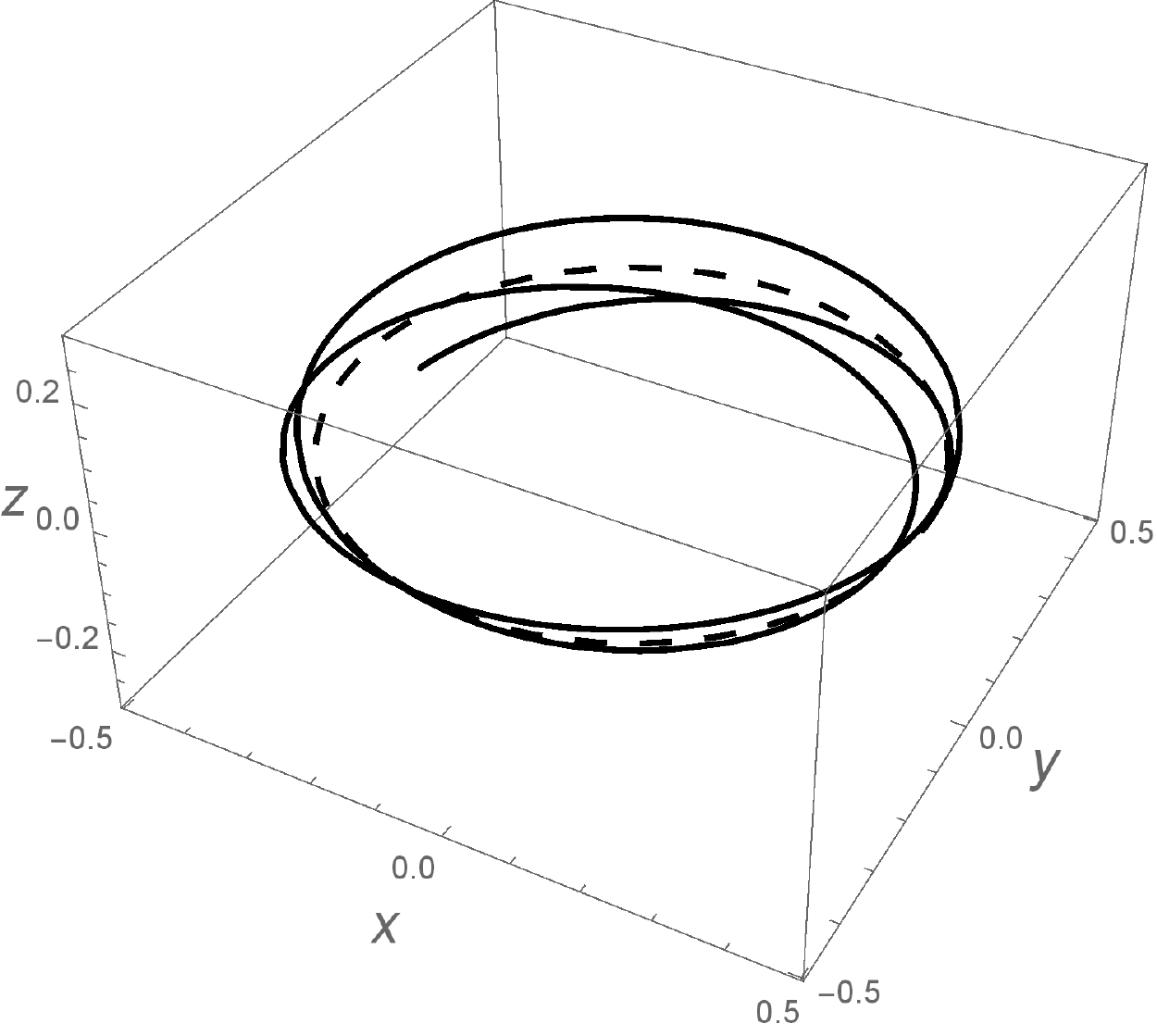}
\caption{\small The circular motion (dashed-line) \\ and perturbed (solid-line) motion
around \\ the $z$ axis. The plot is by values:  $\mu=1$,\\ $\ell=1$, $eg/c=1$, $c\, p_\phi/eg=2.12$, $\rho_0=0.42$,\\ $\dot{\rho}_0=0.05$, 
$z_0=0$, $\dot{z}_0=0.06$, $\phi_0=0$,\\ $\dot{\phi}_0=1.57$}
\end{subfigure}
	\caption{}
\end{figure}

\subsection{Scattering}
The above analysis on the effective potential can be used to study the  scattering processes as well. 
The problem of scattering of an electric charge by a single 
monopole is studied in \cite{boulware,kazama,urru}. 
In the system by one monopole the total angular momentum 
$\mathbf{J}$ of charge and electromagnetic 
fields is conserved, by which it can be shown that the motion of electric charge is entirely 
on a cone with the monopole at its apex \cite{boulware}. 
This fact lets to eliminate the coordinate parallel to the axes of cone, 
leading to an exact solution for the motion projected on the plane perpendicular to the 
axes of cone \cite{boulware}. The relation between the scattering angle $\theta$ 
and the impact parameter $b$ in the reduced two dimensional problem can 
be obtained analytically \cite{boulware}, which can be inserted in the classical relation \cite{goldstein}:
\begin{align}
\frac{d\sigma}{d\Omega}= \frac{b}{\sin\theta} \left|\frac{db}{d\theta}\right|
\end{align} 
For the small impact parameters, where the cone is narrower, 
the scattering angle $\theta$
oscillates about the scattering angle $\pi$ (backward direction) \cite{boulware}. The cross section goes to infinity whenever the oscillating function passes $\theta=\pi$, leading to the 
so-called backward glory phenomena \cite{goldstein}. Also due to 
the vanishing of $d\theta/db$ at the extrema of the oscillating function $\theta(b)$, 
the cross section also diverges at some other impact parameters, leading to the 
so-called rainbow effect \cite{boulware}. 

In the present problem the rotational symmetry is explicitly broken to an axial one
around the axes passing the monopoles, here is taken to be the
$z$-axes. As the consequence, only the $z$ component of the total angular 
momentum, here denoted by $p_\phi$ in (\ref{15}), is conserved. 
Similar to the case with one single monopole, the axial symmetry
around $z$ direction suggests to project the motion on the $xy$-plane, however
in this case the motion is not confined to a cone and numerical solutions are to be 
developed. 
By the analysis based on the effective potential we already know that for 
$0 \leq \frac{c\, p_\phi}{eg} \leq  2.18$ the potential has an attractive part,
beyond it the potential is totally repulsive. It is known that when the potential
has an attractive part the so-called quasi-stable orbiting paths may develop 
infinite values for the polar angle \cite{goldstein}. Further one should expect vanishing 
deflection angle for a finite value of impact parameter for which the attractive part 
of potential cancels out the effect of the repulsive part, leading to the forward glory effect
\cite{goldstein}. Interestingly both of these effects are present for the scattering from
a monopole pair. For the incoming charges with $\mathbf{v}_0=-v_0\, \mathbf{\hat{x}}$ 
in the $xy$-plane, for which $p_\phi=\mu\, b\, v_0$, 
the results of the numerical solutions as the paths in the $xy$-plane 
for different impact parameters are plotted in Fig.~5. 
The paths clearly show both the repulsive and the attracting effects 
on the incoming charges. In fact for appropriate values 
of $p_\phi$ together with sufficient value of $E_0=\mu v_0^2/2$ to 
overcome the maximum of the potential hump (see Figs.~3~\&~4), the charges 
are affected by the attractive part too. 
The values of the polar angles of outgoing charges for different impact parameters 
are plotted in Fig.~6a. The result is quite in accordance with expectations from 
the scattering theory by potentials with attractive part \cite{goldstein}. 
All the incoming angles have the initial polar angle $\phi_0=0$, and 
so those with $\phi_\infty\to\pm\pi$ are to be considered as 
undeflected ones. As usual the large impact parameters $b\to\pm\infty$ 
lead to the undeflected ones at two ends of the plot in Fig.~6a. 
The values of polar angles by three values of impact parameters 
are to be recognized. First is about the impact parameter
at which the deflection angle is $\pi$ (polar angle $\phi_\infty=0$).
This impact parameter is denoted by $b_1$ in Fig.~6a, 
at which the cross section diverges and is known as the 
backward glory effect. The second is about the impact parameter at which 
the repulsive and attractive parts in potential cancel out the effect
of each other, leading to zero deflection ($\phi_\infty=\pi$). 
This impact parameter is denoted by $b_2$ in Fig.~6a, at which the cross section 
is diverging and is known as the forward glory effect.
The third is the impact parameter at which the energy is equal to the 
potential hump and the charge makes a quasi-stable orbiting 
path, leading to infinite polar angle. This impact parameters
is $b_3$ in Fig.~6a. The paths corresponding to the impact parameters
$b_1$, $b_2$ and $b_3$ are plotted in Fig.~6b as paths 1, 2 and 3, respectively. 

\begin{figure}[t]
	\begin{center}
		\includegraphics[scale=0.6]{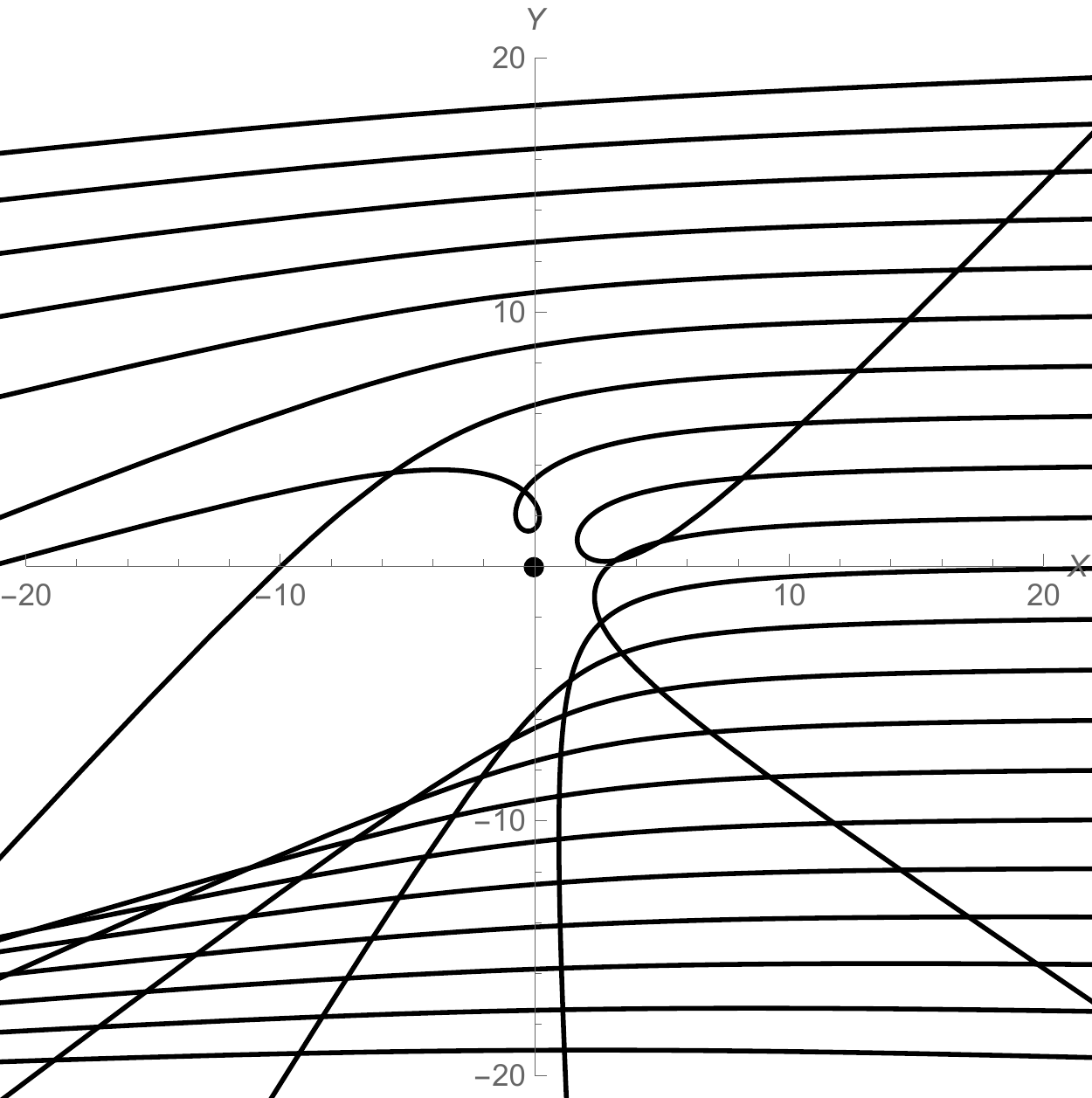}
	\end{center}
	\caption{\small The paths in the $xy$-plane for different impact parameters for
values $\mu=1$, $\ell=1$, $eg/c=2.8$, and $v_0=1.2$.
}
\end{figure}

\begin{figure}[H]
\centering
\begin{subfigure}{.5\textwidth}
\centering
		\includegraphics[width=.9\linewidth]{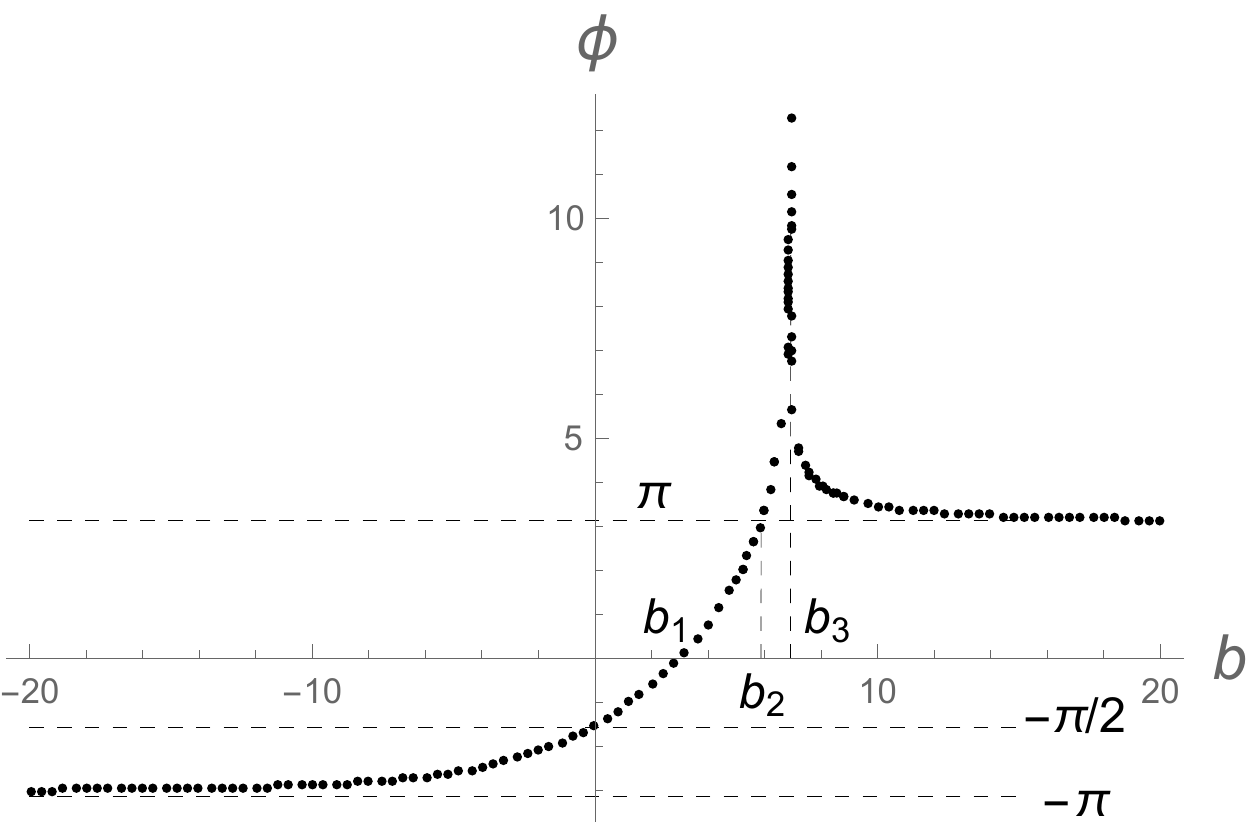}
\caption{\small The polar angle versus the impact \\ parameter 
by the numerical solutions for \\ initial values in Fig.~5. 
The introduced \\ values are  $b_1=2.97$, $b_2=5.88$,
and \\ $b_3=6.92$. }
\end{subfigure}%
\begin{subfigure}{.5\textwidth}
\centering
		\includegraphics[width=.9\linewidth]{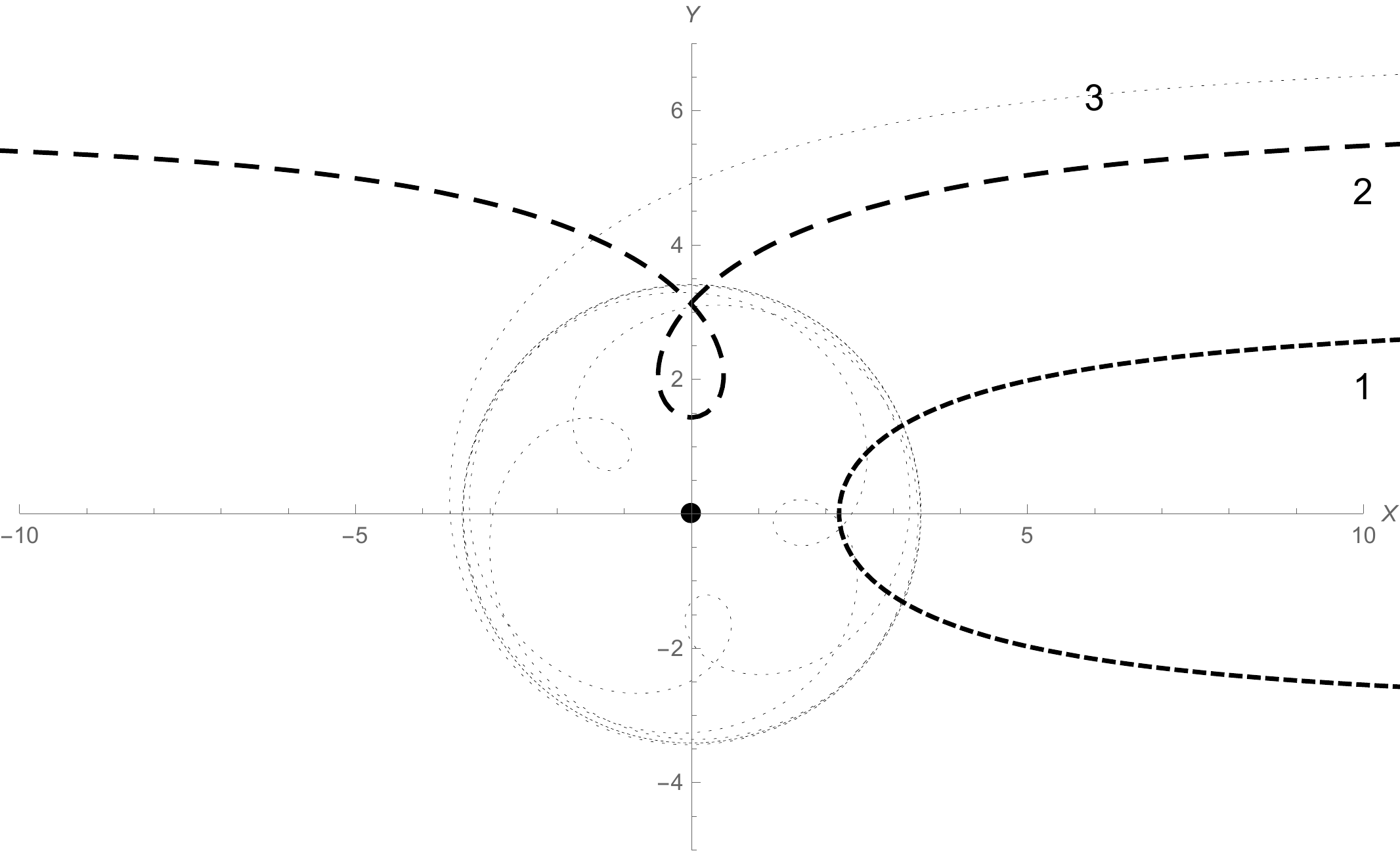}
\caption{\small The three paths 1, 2 and 3 for impact \\ parameters 
$b_1$, $b_2$, and $b_3$, corresponding\\
to backward, forward and quasi-stable \\ paths, respectively.}
\end{subfigure}
	\caption{}
\end{figure}

\section{Quantum Dynamics}

The Hamiltonian operator in the cylindrical coordinates takes the form
\begin{align}\label{31}
H=-\frac{\hbar^2}{2\,\mu} \left(
\frac{1}{\rho}\frac{\partial}{\partial\rho}\rho\frac{\partial}{\partial\rho}
+\frac{\partial^2}{\partial z^2}
\right)+  
 \frac{1}{2\,\mu\,\rho^2}\left(p_\phi - \frac{eg}{c} 
\,\left(\frac{z+\ell}{r_-} - \frac{z-\ell}{r_+}\right)\right)^2
\end{align}
In the quantum theory the quantization conditions read
\begin{align}\label{32}
p_\phi=m\,\hbar,~~~~~~m=0,\pm 1,\pm 2,\cdots 
\\\label{33}
eg/c=q\,\hbar/2,~~~~~~q=0,\pm 1,\pm 2,\cdots
\end{align}
Replaced by the dimensionless coordinates 
$\rho\to \ell\,\rho$ and $z\to \ell\, z$, the Hamiltonian for fixed 
$p_\phi=m\hbar$ takes the form:
\begin{align}\label{34}
H_m=\frac{\hbar^2}{2\,\mu\ell^2} 
\left[-\frac{1}{\rho}\frac{\partial}{\partial\rho}\rho\frac{\partial}{\partial\rho}
-\frac{\partial^2}{\partial z^2}   +  
 \frac{1}{\rho^2}\left( m -\frac{q}{2} 
\,\left(\frac{z+1}{r_-} - \frac{z-1}{r_+}\right)\right)^2\right]
\end{align}
in which $\frac{\hbar^2}{2\,\mu\ell^2} $ has the energy dimension, and 
we have for the dimensionless distances
\begin{align}\label{35}
r_\pm=\sqrt{\rho^2+(z\mp 1)^2} 
\end{align}
The Schrodinger equation for fixed $p_\phi=m\hbar$ then might be represented as
\begin{align}\label{36}
H_m \Psi_m(\rho,z) =\frac{\hbar}{2\,\mu\ell^2}\, E_m \, \Psi_m(\rho,z)
\end{align}
in which $E_m$ is the dimensionless eigen-value of $H_m$.
Since the $\ell$-dependence can be factored out, 
the nature of the eigen-functions is independent of the value of $\ell$. 
It is known that due to the tunneling effect, in presence of a finite height and 
width barrier a true bound-states do not exist in the quantum theory.
In the present problem this simply means that, the charge initially located around the minimum 
of the potential will eventually tunnel to infinity through the barrier. 
The irrelevance of the value of $\ell$ in this respect can be understood simply 
by the tunneling effect as follows. 
It is known that the transmission rate $T$ depends on the barrier height $\Delta V$ and its 
width $\Delta L$ through the relation 
$T\simeq \exp\left(-2\sqrt{\frac{2\mu}{\hbar^2}\,\Delta V}\,\Delta L\right)$. 
For the value of the potential at extrema by condition (\ref{24}) we easily find the finite height:
\begin{align}\label{37}
V_\mathrm{eff}\Big|_{\mathrm{extrema}} & = \frac{2}{\mu}
\left(\frac{eg}{c} \right)^2 \, 
\frac{\ell^2\rho_0^2}{r_0^6}
\cr
& = \frac{2}{\mu}
\left(\frac{eg}{c} \right)^2 \frac{1}{\ell^2}\,
\sin^2\theta_0 \, \cos^4\theta_0
\end{align}
As $\ell$ is the only parameter of length dimension in the 
system, we have 
\begin{align}\label{38}
\left.
\begin{array}{ll}
\Delta V \propto \left(\frac{eg}{c\,\ell}\right)^2 \\
\Delta L \propto \ell
\end{array}
\right\} \to \sqrt{\Delta V}\Delta L \propto eg/c
\end{align}
showing that the transition rate is independent of $\ell$. Further the above
shows that for large enough $eg/c$ the transition rate is so small that the 
approximate bound-states are expected. By (\ref{32}) and (\ref{33}) we have 
\begin{align}\label{39}
\frac{c\,p_\phi}{eg} = \frac{2\,m}{q}
\end{align}
Based on the above, one may try to estimate the approximate bound-state spectrum
in the large $m$ limit with fixed $2\,m/q$, for which the electric charge may
be considered a bound particle by the monopole pair. 
Here we use the variational method of Rayleigh-Ritz, in which one 
starts by a set of basis functions \cite{merz}:
\begin{align}\label{40}
\chi_1,\chi_2, \cdots,\chi_N
\end{align}
In general the basis functions are neither orthogonal nor normalized. 
The variational trial functions are then expanded as
\begin{align}\label{41}
\psi=\sum_{i=1}^N c^i\,\chi_i
\end{align}
The aim is to determine the coefficients $c^i$'s via the variational 
method to minimize the expectation $\langle H \rangle$. The 
metric and the Hamiltonian matrix elements are then defined as \cite{merz}:
\begin{align}\label{42}
H_{ij}&= \langle\chi_i | H |\chi_j\rangle = \int d^3x\, \chi^\star_i H \chi_j
\\\label{43}
g_{ij}&= \langle\chi_i | \chi_j\rangle = \int d^3x\, \chi^\star_i \, \chi_j
\end{align}
Eventually, the eigen-values and eigen-vectors of 
the Hamiltonian is obtained through the system of equations \cite{merz}
\begin{align}\label{44}
&\sum_{j=1}^N H^i_{~j}\,c^j = E\,c^i 
\\\label{45}
&\det \left(H^i_{~j} - E\, \delta^i_{~j} \right) = 0
\end{align}
in which $H^i_{~j}$ is defined by the inverse of the metric $g^{ik}$ \cite{merz}:
\begin{align}\label{46}
H^i_{~j}= \sum_{k=1}^N g^{ik} H_{kj},~~~~~~ 
\sum_{k=1}^N g^{ik} g_{kj}=\delta^i_{~j}
\end{align}
It can be shown that if one adds the level of truncation from $N$ to $N+1$,
the spectrum obtained by (\ref{45}) would remain unchanged or decreased, as 
one expects in a variational approach. In the following we use the above method
to estimate the energy spectrum of the system.

\subsection{$0< \displaystyle{\frac{2m}{q}} <2$: Bound-States in Minimum Valley}
Based on (\ref{20}), the minimum valley is defined by $f(\rho,z)=0$ with
\begin{align}\label{47}
f(\rho,z)= m -\frac{q}{2} 
\,\left(\frac{z+1}{r_-} - \frac{z-1}{r_+}\right)
\end{align}
Due to square nature of the effective potential (\ref{17}), the basis in this case 
are taken to be the combination of the one by Harmonic oscillator,
representing oscillations transverse to the valley, plus the 
standing wave, responsible for oscillations along the valley, 
\begin{align}\label{48}
&\chi_{nn'}(\rho,z)=\exp\!\left[
-\frac{\alpha^2}{2}\left(\frac{f(\rho,z)}{\rho(\rho_{02}-\rho)} \right)^2\right]
\mathrm{H}_n\!\big(\alpha f(\rho,z)\big)\,
\sin(n' \theta)\\
&n= 0,1,2, \cdots, N,~~~~~ n'= 0,1,2, \cdots, N'
\nonumber
\end{align}
in which $\mathrm{H}_n$ is the $n$-th Hermite polynomial, 
$\rho_{02}$ is the radius at the saddle-point, and
$\theta$ is the usual polar angle $\theta= \tan^{-1}(\rho/z)$.
The plots of four members of the basis set are presented in Fig.~7.
In above $\alpha$ is treated as an extra variational parameter by which
we minimize the expectation value. 
The integration domains in (\ref{42}) for the dimensionless 
coordinates are taken to be:
\begin{align}\label{49}
0\leq \rho \leq \rho_{02},~~~~ -1\leq z\leq 1
\end{align}
The denominator $\rho_{02}-\rho$ is inserted
in the exponential to ensure the vanishing of the basis functions
at saddle-point radius. 

\begin{figure}[t]
\begin{subfigure}{.5\textwidth}
\centering
		\includegraphics[width=.7\linewidth]{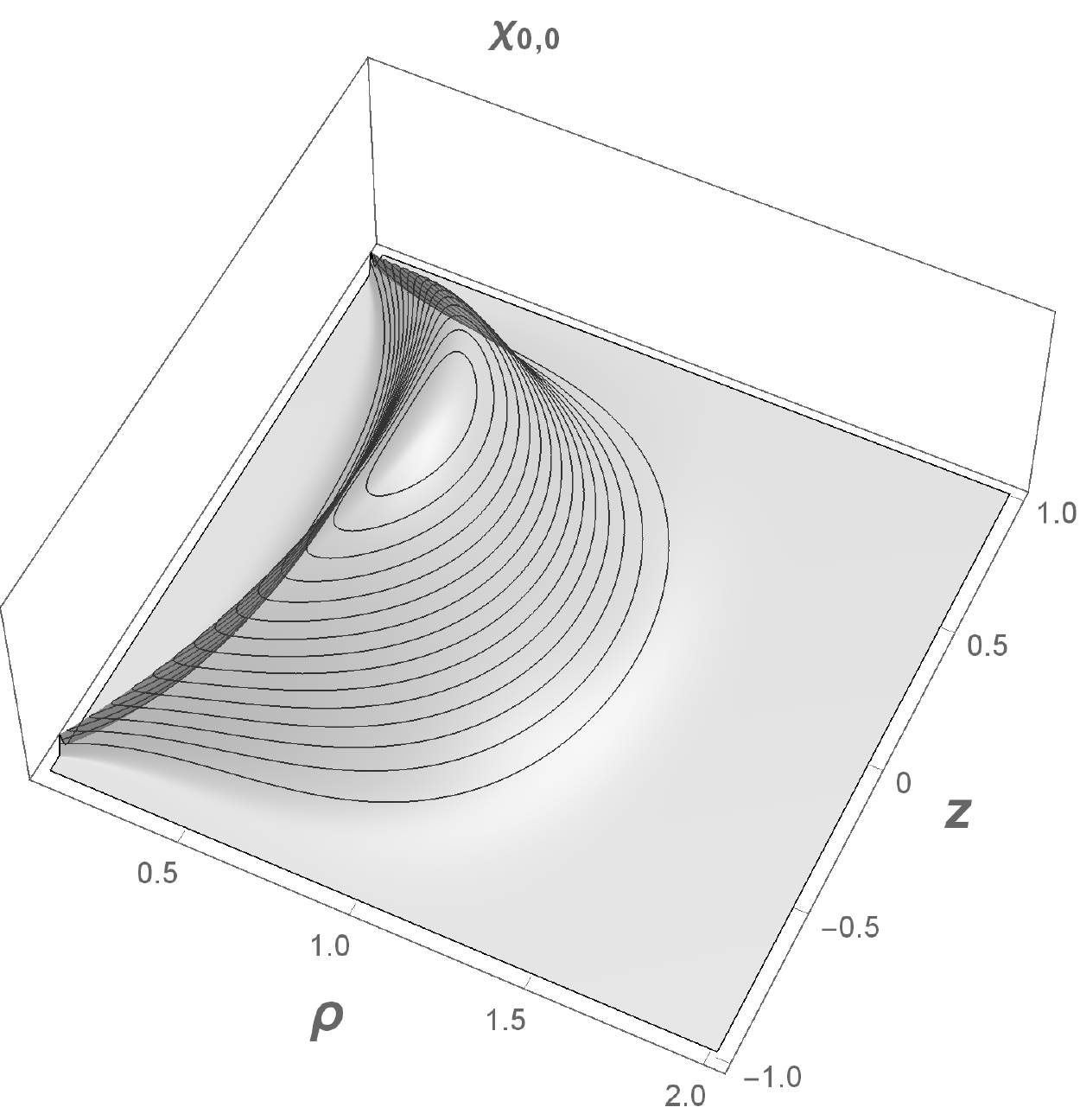}
\end{subfigure}%
\begin{subfigure}{.5\textwidth}
\centering
		\includegraphics[width=.7\linewidth]{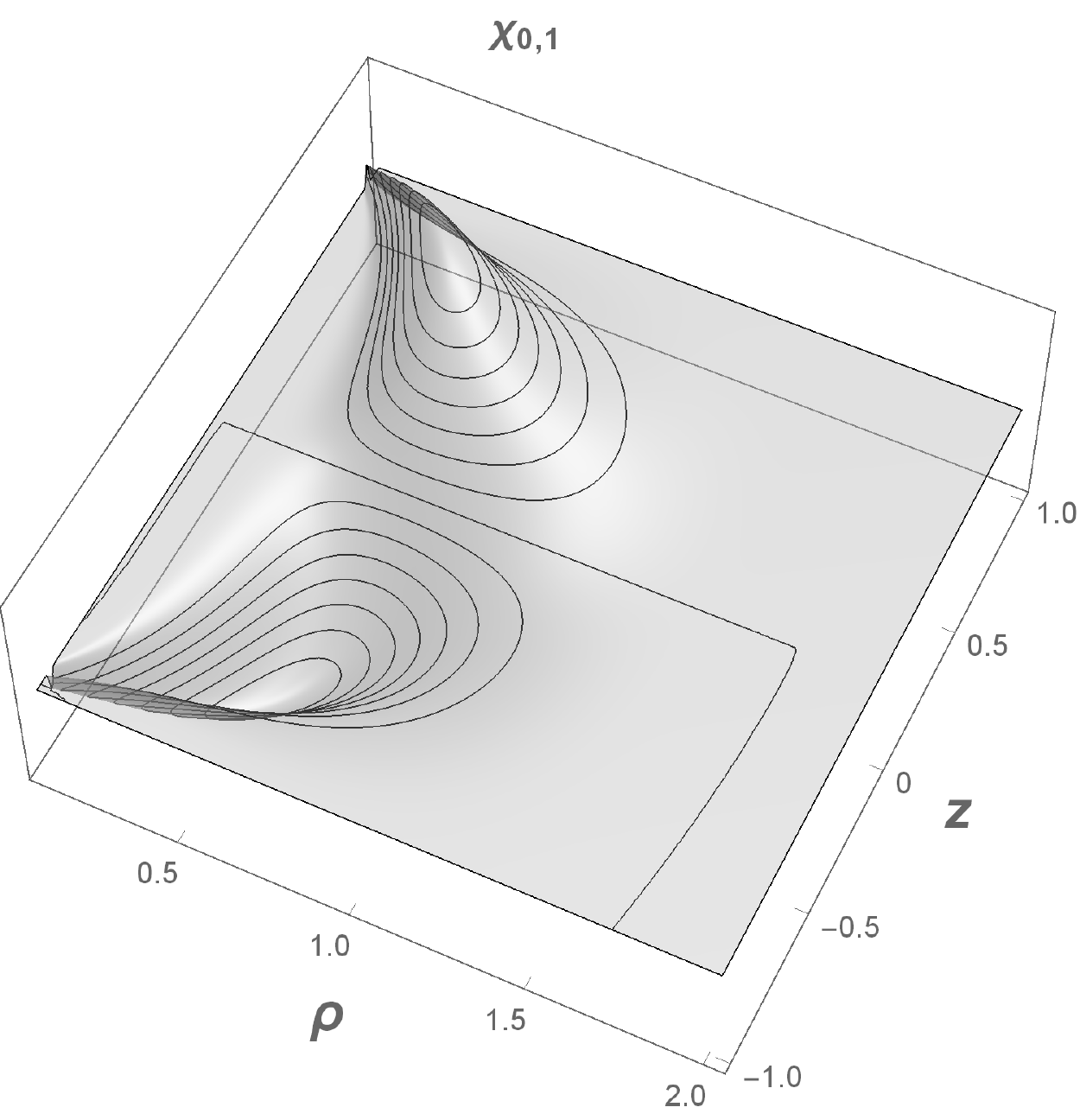}
\end{subfigure}
\begin{subfigure}{.5\textwidth}
\centering
		\includegraphics[width=.7\linewidth]{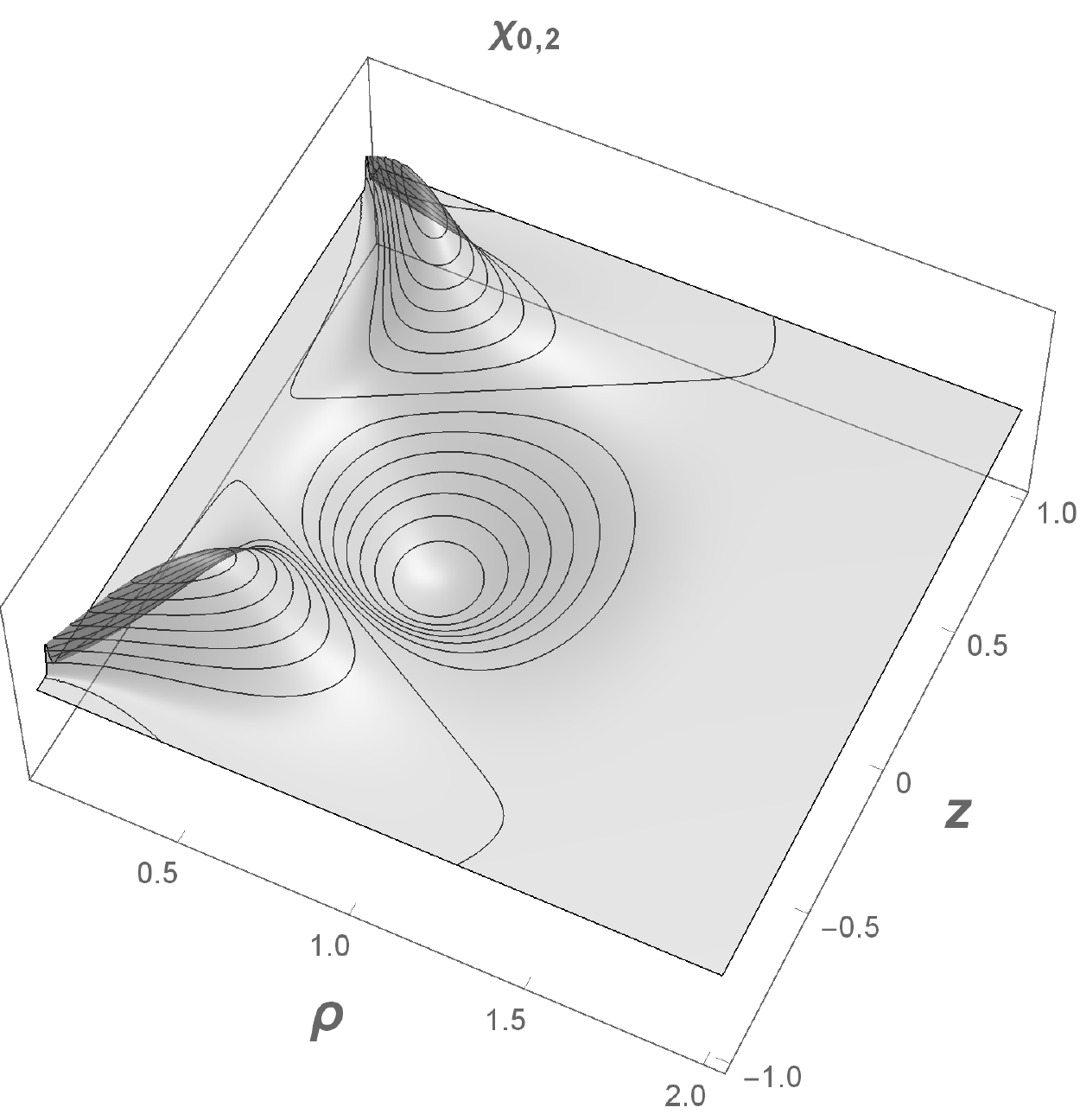}
\end{subfigure}%
\begin{subfigure}{.5\textwidth}
\centering
		\includegraphics[width=.7\linewidth]{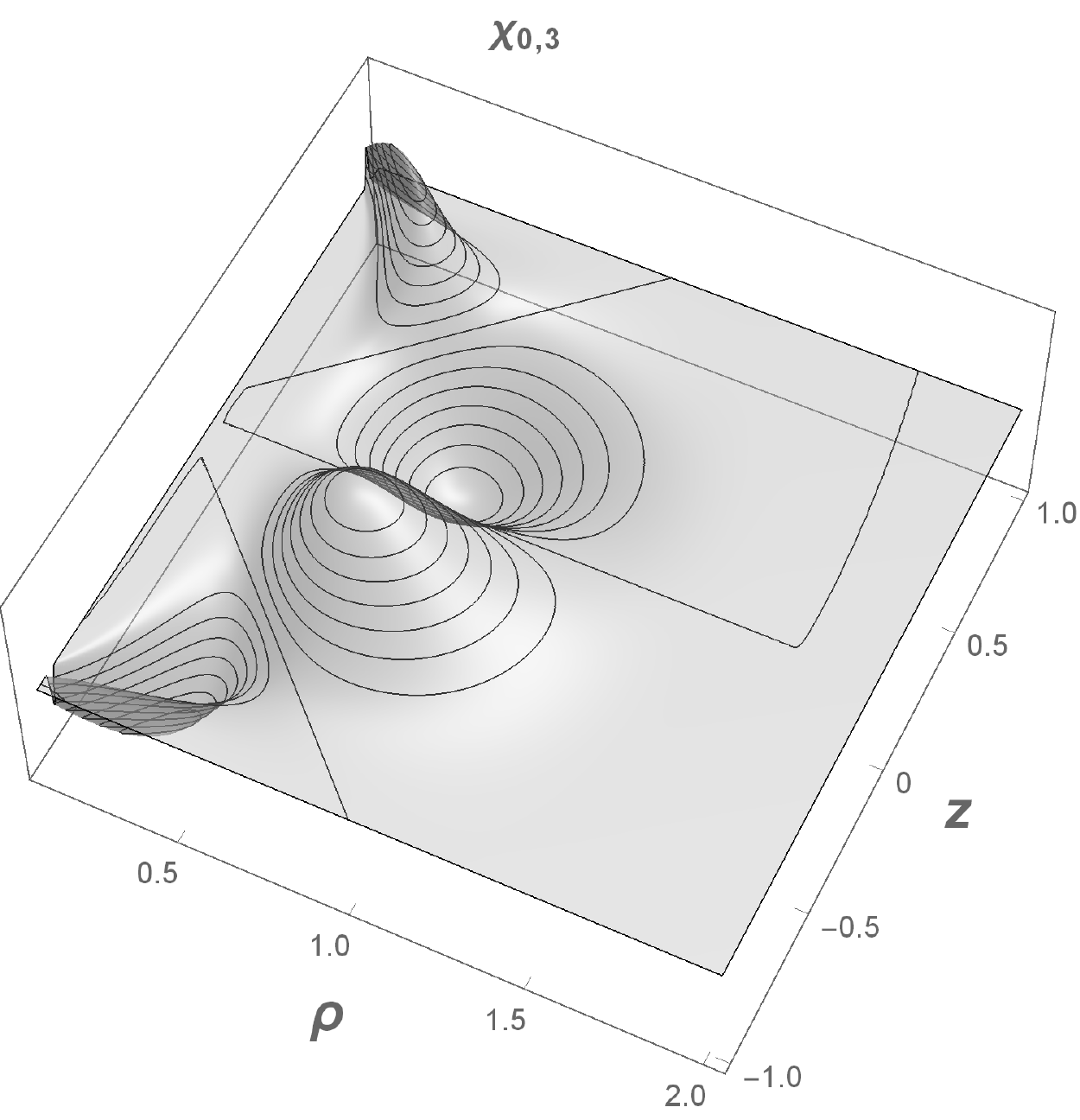}
\end{subfigure}
	\caption{\small Four members of the trial basis functions by (\ref{48}) for 
$m=40$, $q=50$ and $\alpha=0.21$.}
\end{figure}

Based on (\ref{42})-(\ref{46}), 
we find convergence in the lowest two levels by changing the level of 
truncations by the two parameters $N$ and $N'$ from $2$ to $5$; 
the number of basis functions are $(5+1)^2=36$ for the final trial.
For the values:
\begin{align}\label{50}
m=40,~~~~~q=50,~~~~2m/q=1.6
\end{align}
with $V_m=77.8$ as the value of potential at the saddle-point, 
the result by the variational method is as following ($\alpha=0.21$):
\begin{align}\label{51}
\psi_{E_1}(\rho,z) &\simeq 0.95\, \chi_{00} -0.31\, \chi_{02},~~~~~~~E_1=31 
\\\label{52}
\psi_{E_2}(\rho,z) &\simeq 0.89\, \chi_{01} - 0.35\, \chi_{03},~~~~~~~E_2=48
\end{align}
with $E_3=65$ which is comparable to the potential at the saddle-point 
(all energy values are in units $\hbar^2/2\mu\ell^2$). 
The obtained eigen-functions are plotted in Fig~8. 

\begin{figure}[H]
\centering
\begin{subfigure}{.5\textwidth}
\centering
		\includegraphics[width=.9\linewidth]{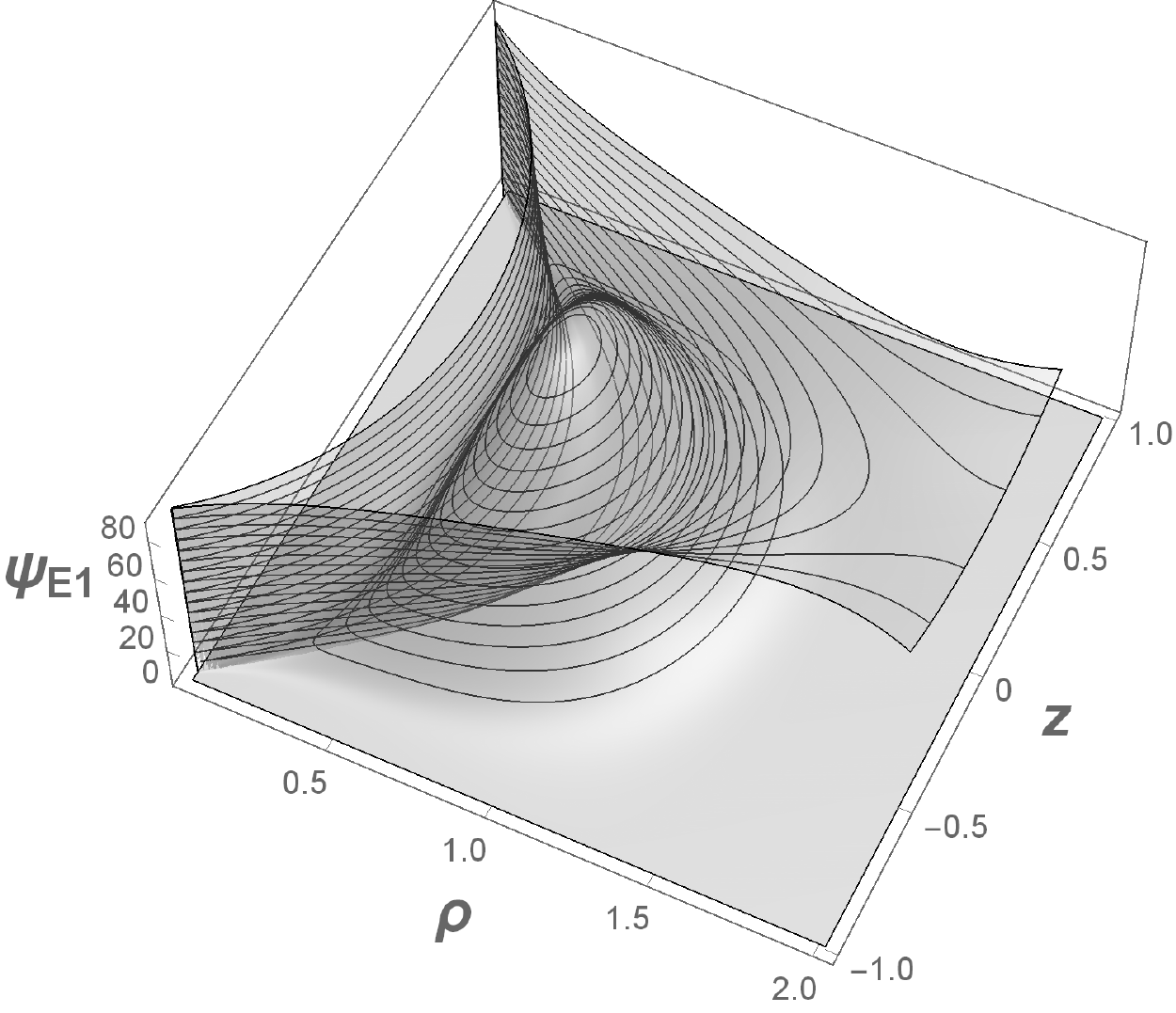}
\end{subfigure}%
\begin{subfigure}{.5\textwidth}
\centering
		\includegraphics[width=.9\linewidth]{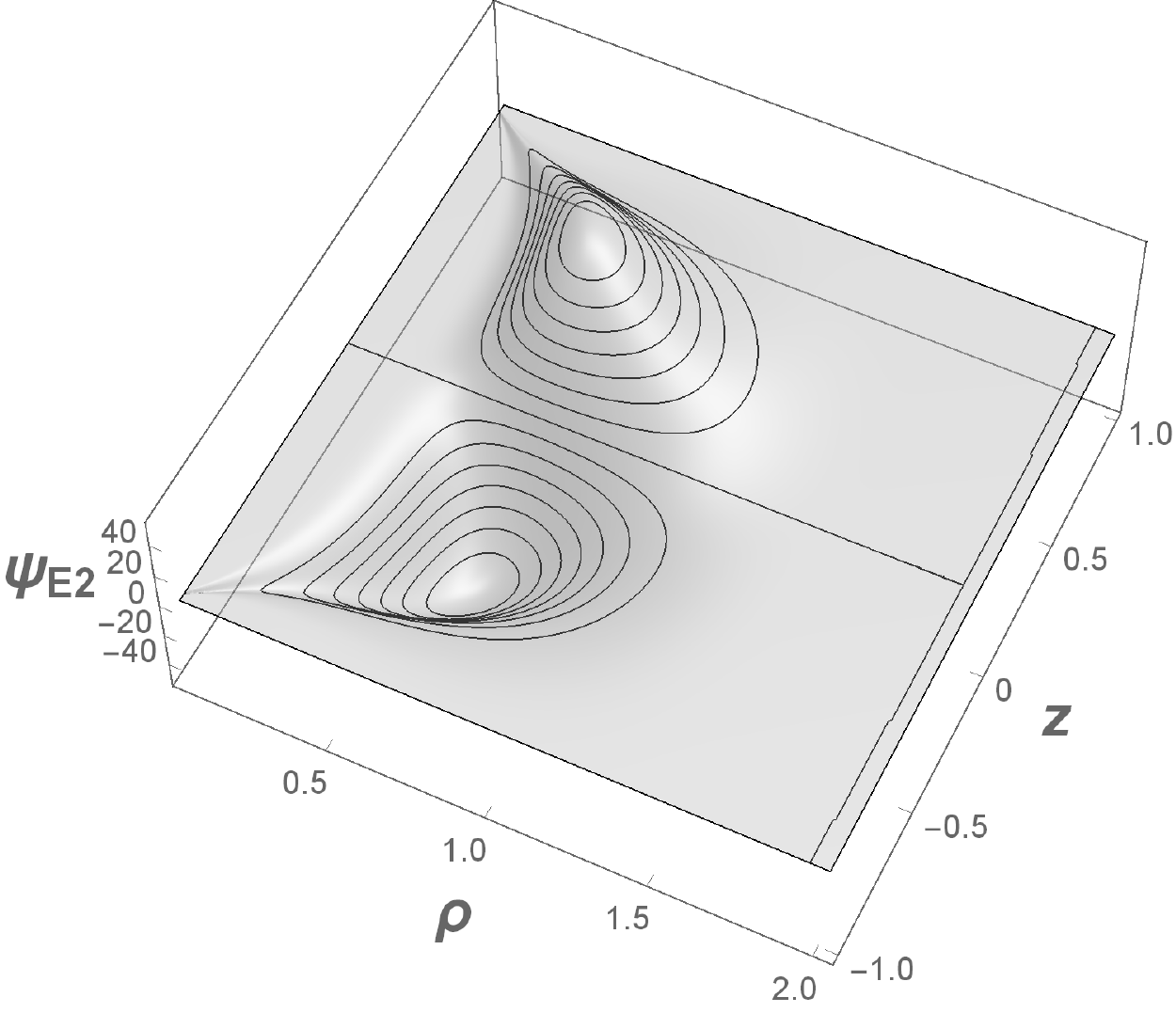}
\end{subfigure}
	\caption{\small The scaled eigen-functions by (\ref{51}) and (\ref{52}). For the 
ground-state (left) the effective potential is also plotted.}
\end{figure}

\subsection{$2 \leq \displaystyle{\frac{2m}{q}} \leq 
\displaystyle{\frac{8\sqrt{2}}{3\sqrt{3}}}$: Bound-States in Minimum Well}
In the case that the minimum makes a well-shape, the trial functions may be 
developed based on the quadratic expansion of the potential around the 
minimum. In particular, using
\begin{align}\label{53}
\frac{\partial^2V_\mathrm{eff} }{\partial\rho^2}\bigg|_{\rho_{01},0}&=\omega^2_\rho
\\\label{54}
\frac{\partial^2V_\mathrm{eff} }{\partial z^2}\bigg|_{\rho_{01},0}&=\omega^2_z
\\\label{55}
\frac{\partial^2V_\mathrm{eff}}{\partial\rho\partial z}\bigg|_{\rho_{01},0}&=0
\end{align}
in which $\rho_{01}$ is the minimum point by (\ref{24}), we may write
\begin{align}\label{56}
V_\mathrm{eff}(\rho,z) = V_\mathrm{eff}(\rho_{01},0)+\frac{1}{2}
\omega^2_\rho (\rho-\rho_{01})^2+\frac{1}{2}\omega^2_z\,z^2 +\cdots
\end{align}
Then one simply takes the trial basis functions by the combination of two 
harmonic oscillators as
\begin{align}\label{57}
\chi_{nn'}(\rho,z)=&\exp\!\left[
-\frac{\alpha^2}{2}\omega_\rho \left(\frac{\rho-\rho_{01}}{\rho(\rho_{02}-\rho)} \right)^2\right]
\mathrm{H}_n\!\big(\alpha \sqrt{\omega_\rho} (\rho-\rho_{01})\big) 
\cr&~\times 
\exp\!\left[
-\frac{\beta^2}{2}\omega_z z^2 \right]
\mathrm{H}_{n'}\big(\beta \sqrt{\omega_z} z \big)
\\\nonumber
&n= 0,1,2, \cdots, N,~~~~~ n'= 0,1,2, \cdots, N'
\end{align}
In Fig.~9 two members of the trial basis functions are plotted.
As a specific example the variational method is applied to the case by values 
\begin{align}\label{58}
m=52,~~~~~q=50,~~~~2m/q=2.08
\end{align}
with $V_\mathrm{min}=187$ and $V_\mathrm{sadd.}=279$, and
\begin{align}\label{59}
\omega^2_\rho=2740,~~~~~\omega^2_z=1022
\end{align}
The result by the variational method is as following ($\alpha=0.21$
and $\beta=0.91$):
\begin{align}\label{60}
\psi_{E_1}(\rho,z) &\simeq 0.94\, \chi_{00} +0.33\, \chi_{10},~~~~~~~E_1=250 
\end{align}
with $E_2=300$ which is more than the potential at saddle-point, and so can 
not present a bound-state. The corresponding wave-function is plotted in Fig.~10.

\begin{figure}[t]
\centering
\begin{subfigure}{.5\textwidth}
\centering
		\includegraphics[width=.9\linewidth]{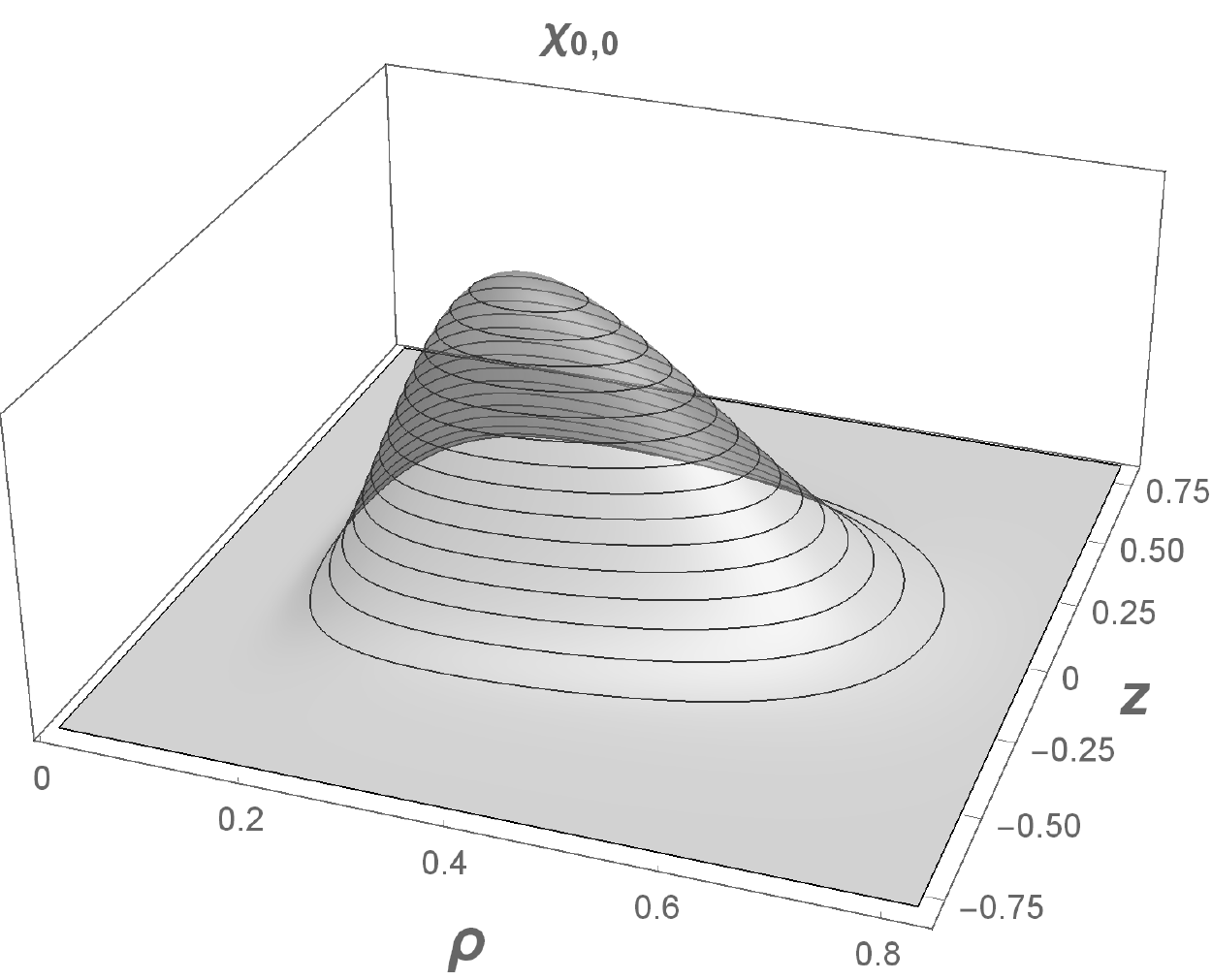}
\end{subfigure}%
\begin{subfigure}{.5\textwidth}
\centering
		\includegraphics[width=.9\linewidth]{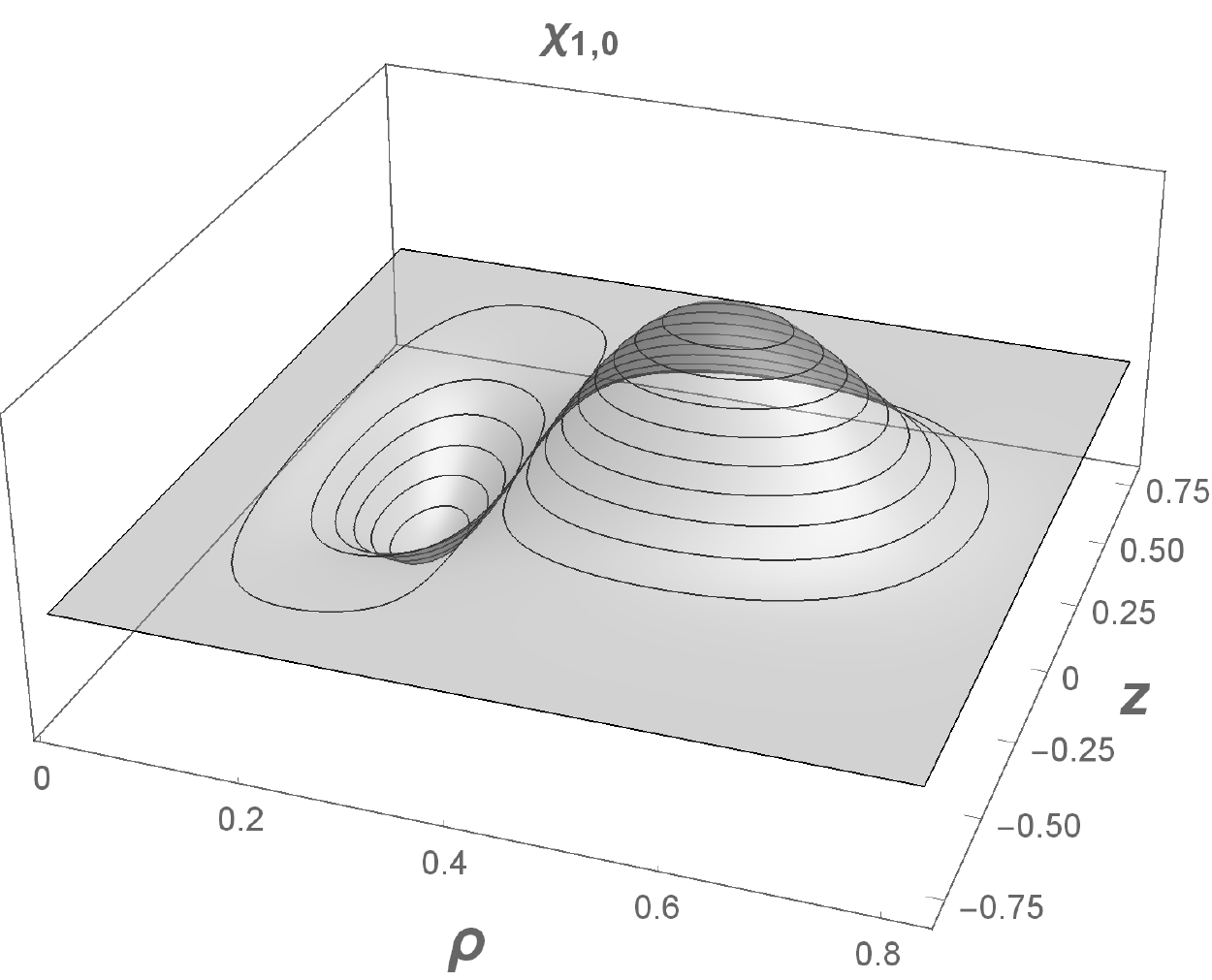}
\end{subfigure}
	\caption{\small Two members of the trial basis functions by (\ref{57}) for 
$m=52$, $q=50$, $\alpha=0.21$ and $\beta=0.91$.}
\end{figure}

\begin{figure}[H]
	\begin{center}
		\includegraphics[scale=0.7]{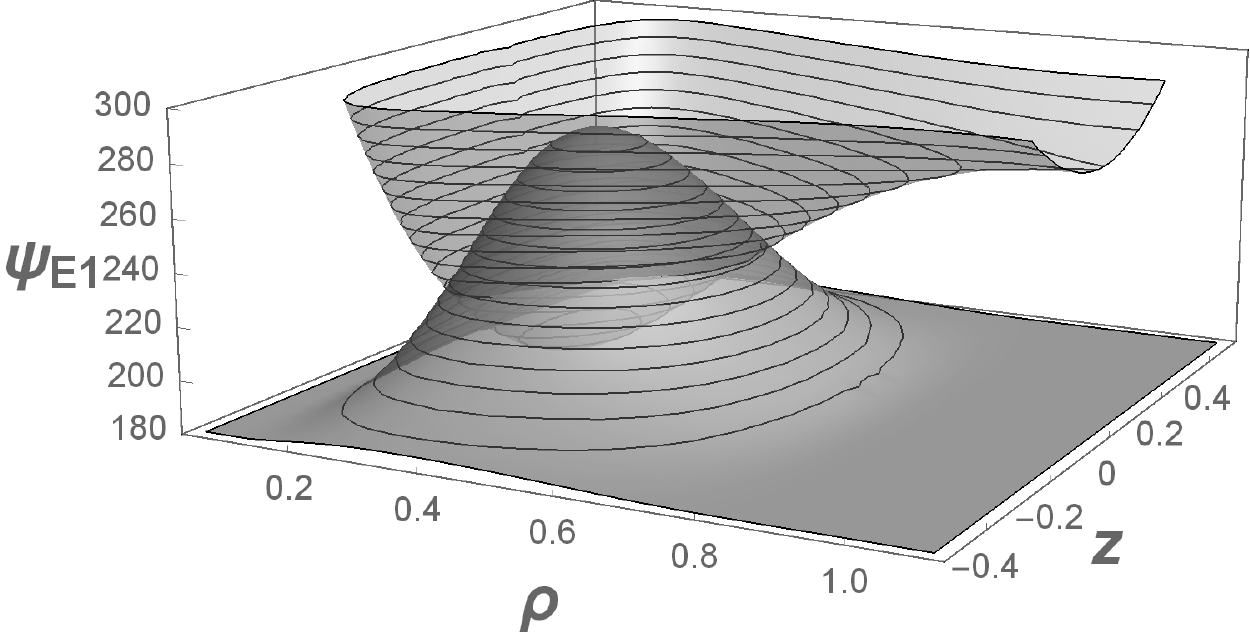}
	\end{center}
	\caption{\small The scaled eigen-function by (\ref{60}) together with 
the effective potential.  }
\end{figure}

\section{Conclusion and Discussion}
The classical and quantum mechanics of  
the electric charge $e$ in presence of a fixed $\pm g$ pair of magnetic poles 
is studied. In the Hamiltonian formulation of the system the effective potential 
$V_\mathrm{eff}(\rho,z)$ describes the dynamics. 
It is found that depending on the ratio of the angular momentum $p_\phi$
and the combination $eg/c$ the potential may develop extrema. 
For $0\leq cp_\phi/eg \leq 2$ the potential 
consists a zero valley minimum along a curve connecting the poles. For 
$2\leq cp_\phi/eg \lesssim 2.18$ the potential consists a finite depth well, 
and for higher $cp_\phi/eg $ no extrema is present. The minimum valley or well are 
separated from $V\to 0$ tail at infinity by a finite height barrier at a saddle-point. 
In the classical regime,  the case with a minimum valley  
develop the bound motions along a curve between two poles,
and the case with the minimum well may develop
the motion around the axis passing two poles. When no extrema
in the effective potential naturally no bound-state is expected.
In the quantum mechanical setup, due to the tunneling effect through 
the finite height barrier, the classical bound motions turn to the so-called 
quasi bound-states. However, for large values of $eg/c$, for which 
it is shown that the height of barrier at saddle-point is large, the states 
may be treated approximately as true bound-states, for which the energy 
eigen-values may be evaluated. For these cases by the variational Rayleigh-Ritz 
method the energy as well as eigen-functions are obtained. 

One natural extension of the present problem is to study the statistical 
mechanics of the electric charges in presence of the monopole pair. 
In such an extension one may include the dynamics of the fields as well. 
It is of great importance to study the mutual effect of moving charges and the 
magnetic fields, and its consequence on the phase structure of the system. 
This last extension is specially important considering the dual picture of the 
present problem, in which a gas of monopoles is being considered in presence 
of two opposite electric charges. As mentioned earlier, according to
the confinement mechanism based on the dual picture 
of the so-called Meissner effect in type-II superconductors, 
it is the circulating motion of the monopoles that prevents the 
spreading the electric fluxes, leading to the confinement of the 
electric charges \cite{nambu,mand,thooft}. 
Based on the results presented here, by considering only the dynamics of charges with given electric field-lines, the system does not develop a bound-state, but only approximate or quasi bound-state. 
It means that for having a successful explanation of the confinement based on the above mechanism, one has to include the dynamics of fields as well. 

\appendix

\section{Extrema by $\partial F = 0$}
The effective potential has the form
\begin{align}\label{61}
V_\mathrm{eff}(\rho,z) &= \frac{1}{2\,\mu}\frac{1}{\rho^2}\left(p_\phi - \frac{eg}{c} 
\,h(\rho,z)\right)^2
\end{align}
in which 
\begin{align}\label{62}
h(\rho,z)=\frac{z+\ell}{r_-} - \frac{z-\ell}{r_+}
\end{align}
By the condition $\partial_\rho V_\mathrm{eff}=0$ we have
\begin{align}\label{63}
&h(\rho,z)=0 \\\label{64}
&p_\phi - \frac{eg}{c}  \left(h+\rho\,\frac{\partial h}{\partial \rho}\right)=0,~~~~
\mathrm{for}~~~~z=0,~\rho=\rho_0
\end{align}
The first in above simply defines the zero-valley minimum. The second
leads to 
\begin{align}\label{65}
p_\phi=2\frac{eg}{c}\frac{\ell}{r_0}\left(1+\frac{\rho_0^2}{r_0^2}\right)
\end{align}
with $r_0=\sqrt{\rho_0^2+\ell^2}$. The above in terms of the angle
$\tan\theta_0=\rho_0/\ell$ come to the form of (\ref{24})
\begin{align}\label{66}
\frac{c\,p_\phi}{eg}=2\cos\theta_0 (1+\sin^2\theta_0)
\end{align}
The right-hand side in above is plotted in Fig.~2, with its extrema
coming from vanishing derivative
\begin{align}\label{67}
\sin\theta_0(1-3\sin^2\theta_0)=0~\to~
\begin{cases}
\theta_0=0\cr
\theta_c=\sin^{-1}\frac{1}{\sqrt{3}}
\end{cases}
\end{align}
The second solution in above corresponds the absolute maximum in Fig.~2
with
\begin{align}\label{68}
\frac{c\,p_\phi}{eg}=2\cos\theta_c (1+\sin^2\theta_c)=
\frac{8\sqrt{2}}{3\sqrt{3}}
\end{align}
The case with $cp\,\phi/eg=2$
simply has the below solution corresponding to (\ref{29})
\begin{align}\label{69}
\cos\theta_m=\frac{-1+\sqrt{5}}{2}
\end{align}

\vspace{2mm}
\textbf{Acknowledgement}: 
This work is supported by the Research Council of the Alzahra University.


\end{document}